# A low-cost, user-friendly rheo-optical compression assay to measure mechanical properties of cell spheroids in standard cell culture plates

Rosalia Ferraro[1,2], Sergio Caserta[1,2,*], Stefano Guido[1,2]

[1]DICMaPI, Università di Napoli Federico II, P.le V. Tecchio 80, 80125 Napoli, Italy

[2]CEINGE Advanced Biotechnologies, Via Gaetano Salvatore, 486, 80131 Napoli, Italy

Contact information: rosalia.ferraro@unina.it (0000-0003-1443-4160), sergio.caserta@unina.it (0000-0002-4400-0059), stefano.guido@unina.it (0000-0002-9247-6852)

*Corresponding author: S.C.

# Abstract

The mechanical characterization of cell spheroids, one of the most widely used 3D biology models *in vitro*, is a hotspot of current research on the role played by the mechanical response of cells and tissues. The techniques proposed so far in the literature, while providing important scientific insights, require a specialized equipment and technical skills which are not usually available in cell biology facilities. Here, we present an innovative rheo-optical compression assay based on microscopy glass coverslips as the load applied to cell spheroids in standard cell culture plates and on image acquisition with an optical microscope or even a smartphone equipped with adequate magnification lenses. Mechanical properties can be simply obtained by correlating the applied load to the deformation of cell spheroids measured by image analysis. The low-cost, user-friendly features of the proposed technique can boost mechanobiology research making it easily affordable to any biomedical lab equipped with cell culture facilities.

# 1. Introduction

Cell spheroids, defined as self-assembled 3D cell aggregates *in vitro*[1], are a widespread model of tissue microenvironment used to recapitulate cell-cell and cell-extracellular matrix (ECM) interactions in several physio-pathological processes[2,3]. Due to their structural and functional similarities with *in vivo* tumours[4-6], spheroids have emerged as a valuable tool in cancer research. In fact, being characterized by the presence of chemical gradients (i.e., oxygen, nutrients, and catabolites), cellular spheroids mimic the heterogeneous structure of tumours and generally consist of three regions[7]: an outer shell with actively proliferating cells, a middle region with quiescent cells, and a nutrient deprived zone (necrotic core)[8] in the centre, due to lack of nutrients and accumulation of waste induced by diffusion limitation[9,10].

Emerging applications of cell spheroids are also found in tissue engineering as constructs for implantable materials and as building blocks in bioprinting of complex 3D structures[11]. In all these applications, the mechanical behaviour[12] of cell spheroids, in concert with the biochemical cell machinery, plays a key role not only in physiological processes, such as development and regeneration, but also in disease[13], e.g., being a critical factor in cancer progression and metastasis[14-16].

Historically, the interest in cell spheroids stemmed from the study of morphogenesis thanks to the seminal work of Wilson[17] and Holtfreter[18] and other authors[19], which inspired Steinberg et al. in developing the differential adhesion hypothesis[20] (later modified as the differential interfacial tension hypothesis[21,22]), based on the analogy with phase separation of immiscible fluids, and the tissue surface tensiometer to measure the apparent tissue surface tension (ATST)[23]. In fact, both emulsions and cellular spheroids are characterized by attractive interactions, and the presence of an equivalent surface tension can be envisaged to explain the resistance of spheroids to deformation and coalescence[24].

In most of these works, cell spheroids are compressed in a parallel plate apparatus and the stress relaxation (annealing) data at times in the order of minutes are fit to theoretical models[25,26] to extract the ATST and other mechanical parameters. The proposed picture is that, following an initial elastic deformation, cell rearrangement at longer times could be described in terms of the ATST.

Further experimental advances to characterize cell spheroids mechanical behaviour, such as the Young's modulus and characteristic relaxation times, include other techniques, such as isotropic compression by osmotic pressure[27,28], atomic force microscopy (AFM)[29,30], micro-indentation[31], tweezers[32], cavitation rheology[33,34] and elastography[35,36]. Each of these methods has its own strengths and limitations, and their results, in term of Young's modulus (E), can vary, depending on the specific technique used, as shown in **Supplementary Table 1** for different immortalized cell lines and human biopsies.

Additionally, mechanical properties are also dependent on size[37] and growth conditions of the cell spheroids[38], showing a time-dependent viscoelastic behaviour, influenced by factors such as cell-cell adhesion and extracellular matrix production.

Although the aforementioned techniques have provided important scientific insights on the mechanical behaviour of cell spheroids, they are based on specialized equipment and technical skills which are not usually available in cell biology facilities. Simple, more accessible experimental methods to measure the mechanical properties of cell spheroids are highly needed. Ideally, such methods should be compatible with standard cell culture techniques and rather inexpensive to allow the high-throughput scale which is often required when addressing biological problems. The need to test large number of samples to measure a few experimental parameters of simple interpretation in a standardized way is especially important in studying disease, where high variability is expected and tools allowing to correlate data are quite important.

Here, we propose an innovative rheo-optical compression assay based on simple materials and equipment usually available in a biology lab. Our approach is based on the combination of a constant stress test (creep) by a using a microscope coverslip to apply the load and data fitting of the resulting deformation with rheological models[37] to extract mechanical parameters. The assay is first validated on agarose gel particles and then carried out on two different cell lines to exemplify its application.

## 2. Results

### 2.1. Standard microscope coverslips can be used to compress spherical gel particles in culture plates under well controlled conditions while imaging the resulting deformation

Our main objective was to develop a methodology allowing one to perform a quantitative mechanical characterization of cell spheroids by using standard equipment easily available in cell culture facilities. Preliminary tests have been carried out on spherical particles made of agarose gel, whose preparation is illustrated in **Figure 1** and described more in detail in Materials and Methods (section **4.1**). Briefly, a droplet of a hot 2.5% agarose solution in water is gelled by immersing it in silicone oil at room temperature. Two images of the so obtained agarose gel particle in the side and top view, the former taken by a smartphone with additional optics and the latter by a light microscope, are shown in **Figure 1a (left)**, where the red colour is due to the addition of Rhodamine B to the agarose solution. The agarose gel particle is then transferred into a cell culture plate filled with silicone oil. A microscope coverslip is placed on the agarose gel particle by using a 3D printed frame (see **Figure 1b**) as a positioning guide (see **Supplementary Video 1**). Under such a mechanical load, the agarose gel particle is compressed, as illustrated by the side and top view images (recorded once equilibrium has been reached), which are included in **Figure 1a (right)** as well. The two images show that the agarose gel particle is deformed with respect to the initial spherical shape. The deformation can be described in terms of the distance R' between particle centre and lateral surface with respect to the initial radius R (see **Figure 1 (right)**). Another measure of particle deformation is given by the vertical displacement δ with respect to R (see **Figure 1 (right)**). By placing more coverslips, a further compression is imposed, and the resulting deformation can be obtained by analysing the shape of the compressed agarose gel particle. For the sake of simplicity, only top view images are taken in the experiments (by light microscopy).

## 2.2. Gel particle compression allows to measure the gel elastic modulus in good agreement with classical shear rotational rheometry

The rheo-optical compression test previously described is based on the progressive application of several coverslips, one by one, on the agarose particle by means of the 3D printed guide shown in **Figure 1b**, each time waiting for equilibrium conditions to be reached (which requires a time scale of a few minutes, as shown in the next subsection). In **Figure 2a**, representative phase-contrast microscopy images of agarose particles compressed under different loads are reported. The value of the engineering stress σ, which is shown on top of each image, is calculated as $\sigma = F_N/\pi R^2$ where $F_N$ is the weight of the glass coverslips minus the Archimedean lift, and R is the initial radius of the uncompressed agarose particle. From left to right the agarose particle is subject to stresses ranging from 0 up to 3.02·10⁴ Pa, the latter corresponding to a load of 5 glass coverslips. The images show that both the contact area and the distance R' between the centre and the lateral surface of the deformed agarose particle increase with the applied load, although the increase of R' relative to R is rather small. The penetration depth δ, i.e., the vertical displacement of the coverslip, can be calculated by volume conservation. If the lateral profile of the deformed agarose particle is taken as a circular arc (as suggested by the side view of **Figure 1a right**), the penetration depth can be obtained from the equation[39]:

$$\frac{4}{3}\pi R^3 = 2\pi R'^2(R-\delta) - \frac{2}{3}\pi(R-\delta)^3 \qquad (1)$$

(a similar value is found by assuming an elliptical shape of the lateral profile, see Materials and Methods). From the so obtained value of $\delta$, the vertical deformation $\varepsilon$ can be calculated as $\delta/R$.

The plot of the compression stress $\sigma$ *vs* the deformation $\varepsilon$, shown in **Figure 2b**, reveals a linear correlation between these two quantities. Hence, by a linear fitting of the data according to Hooke's law, a value of the Young modulus E of $8.4 \cdot 10^4$ Pa can be obtained.

The next step was to validate the value of the Young modulus obtained by the gel particle compression test with data from independent, well-established experimental methods. We used two techniques based on a rotational rheometer with a parallel plate configuration: an unconfined compression test (**Figure 2c**) and an oscillatory shear test in the linear viscoelastic regime (**Figure 2d**). In the former test a sample of the agarose solution is loaded between the plates and let to gel at room temperature. The upper plate is then brought to a lower position and the compression force $F_N$ is measured by the normal force transducer of the rheometer. The compression stress $\sigma$ is calculated as the measured force divided by the area of the plates, while the strain $\varepsilon$ is obtained as the ratio of the current and initial values of the gap between the plates. The so determined stress-strain data are plotted in **Figure 2c** and show a linear trend in the range investigated. The slope of the linear fit provides a value of the Young modulus of $2.9 \cdot 10^5$ Pa, which is in reasonable agreement with the value obtained from the compression test of the agarose gel particle (the difference could be partly attributed to the effect of the interfacial tension acting on the lateral surface of the gel between the two plates).

In the oscillatory shear test a sample of the agarose solution is allowed to gel between the parallel plates as before. The upper plate is then subjected to an oscillatory motion at low amplitude to ensure that the sample is tested in the linear viscoelastic regime. The results of the test are given in terms of the elastic and viscous moduli, G' and G", respectively, as a function of the frequency ω, as shown in **Figure 2d**. As expected in the case of a solid-like gel material, G′ is always higher than G″ and the two moduli are almost parallel to each other in the entire range of frequency investigated. The same measurements were taken both before (triangle up) and after (triangle down) the sample compression test. No significant difference between the two sets of data can be observed for the loss modulus $G''$ (orange symbols). On the contrary, G′ (gray symbols) increased by a factor ~1.25, showing a slight compressional stress stiffening behaviour, as reported in our previous work[12]. In the linear viscoelastic regime, the elastic modulus E can be calculated from the shear modulus G from the equation E = 2(1 + ν) G, where ν is the Poisson's ratio, which was taken as equal to 0.5. By taking $G = \sqrt{G'^2 + G''^2} = 6.7 \cdot 10^4$ Pa from the data of **Figure 2d,** a value of E = 2.0·10$^5$ Pa can be estimated, which is once again in good agreement with the value of E from our compression test of the agarose gel particle.

It can be concluded that the compression assay on a gel particle provides a good estimate of the elastic modulus, and this result suggests that the same technique can be applied to cell spheroids as well, which is the focus of this work. Before going to the next section, which will be addressed to the compression test of cell spheroids, it is interesting to notice in **Figure 2d** that the agarose gel used in this work exhibits a non-negligible viscous modulus, although much smaller than the elastic one. The presence of a viscous component in the rheological behaviour can be linked to the transient response following the application of a mechanical load (see next subsection).

## 2.3. The same rheo-optical compression test can be used to probe the mechanical behaviour of cell spheroids in transient and steady conditions

The cellular spheroids for the rheo-optical compression test were prepared according to the method described in section **4.2**. As schematically showed in **Figure 3a**, the cell spheroids were transferred by using a pipette in a multiwell plate pre-filled with standard culture medium. The 3D printed insert was then used to immerse the coverslips in the culture medium, thus compressing the spheroid, and the multiwell plate was placed on the motorized stage of an inverted microscope. Top view microscopy images of a representative spheroid (made of NIH/3T3 cells) are shown with and without the applied stress ($\sigma = 0$ Pa and $\sigma > 0$ Pa) in the bottom of **Figure 3b**. It can be noticed that, at variance with the compressed agarose gel particles, the cell spheroids are quite flattened right after the application of the first coverslip, thus taking an essentially cylindrical shape. Hence, the vertical deformation $\varepsilon = \delta/R$ was calculated from volume conservation between the initial spherical shape of radius R ($V_0 = \frac{4}{3}\pi R^3$) and the compressed cylindrical shape ($V = 2\pi R'^2 (R - \delta)$). By measuring R and R' (see **Figure 3b**) from image analysis and setting $V = V_0$, the value of $\delta$ can be easily obtained. The same procedure can be applied to study the transient deformation of the cell spheroid following the application of a coverslip until a steady state is reached. In all the experiments the stress was calculated by dividing the weight of the applied coverslips (minus the Archimedean lift) by the area of the compressed cell spheroid.

## 2.4. Steady state stress-strain data of cell spheroids under compression provide values of the elastic moduli corresponding to different cell lines

Representative phase-contrast microscopy images showing the morphological response of NIH/3T3 and PANC-1 spheroids at four different values of σ (from 0 to ~ $4 \cdot 10^4$ Pa), corresponding to an increasing number of coverslips as the applied load, are reported in **Figure 4a** and **c**, respectively. The images show that spheroids of both cell lines get progressively more deformed by increasing the applied load, the more so for the PANC-1 spheroids, which is likely related to their tumour phenotype. A quantitative representation of applied load vs measured deformation is given by the stress-strain data in **Figure 4b** and **d**, where data of a single representative spheroid are presented in the left plot and the average small deformation behaviour of at least 15 spheroids are shown in the right plot. The single spheroid data (left plot) of both cell lines exhibit a linear trend up to a value $\varepsilon^*$ around 1 (region highlighted in yellow), followed by a sharp rise at larger deformations.

As done in the analysis of the agarose gel particles, we have averaged the data from different spheroids at $\varepsilon < \varepsilon^*$ and then fit a linear function to the so obtained results with the elastic modulus E as the only fitting parameter. The so calculated values of E are $(5.4 \pm 0.7) \cdot 10^4$ Pa for NIH/3T3 and $(1.9 \pm 0.3) \cdot 10^3$ Pa for PANC-1 spheroids. These values fall within the range of the elastic modulus of spheroids from different cell lines reported in the literature by using several techniques (see **Supplementary Table 1**). Although our experimental results are still quite preliminary and just meant to exemplify the application of the compression test developed in this work, it can be noticed that the value of E for the PANC-1 tumour cell line is much lower than the one for the NIH/3T3 cell line. This result is in line with the finding that tumour cell spheroids are generally softer than spheroids from healthy cell lines, such as tumour cells are apparently softer than their healthy counterparts[13]. Other variables can also affect the mechanical properties of cell spheroids, such as the cultivation time, the spheroid size and the extent and composition of the extracellular matrix around the cells of a spheroid. We believe that a quantitative investigation of the effect of such variables is one of the main outcomes that can be obtained by a systematic application of the compression test here proposed. A detailed study of the mechanical behaviour of cell spheroids as a function of the relevant variables is however outside the scope of this work.

### 2.5. Steady state and transient stress-strain data of compressed cell spheroids can be used to test the application of rheological viscoelastic models

Apart from the measurement of the elastic modulus in the linear regime, another possible use of the compression test is to evaluate the application of rheological models to describe the full mechanical response of cell spheroids, including the large deformations region at $\varepsilon > 1$. In such region, the steep rise of stress at increasing strain is similar to the mechanical behaviour of polymeric foams under uniaxial compression[40]. An analogy between foams and cell spheroids has been suggested in the literature[41] based on the observation that both materials are characterized by packings of soft particles (bubbles in foams and cells in spheroids).

Here, among the rheological models available in literature[42], we have used as an example a phenomenological model, which has been developed to describe the mechanical behaviour of foams[43]. The model, schematized in **Figure 5a**, is characterized by three elements in a parallel combination: a linear spring with an elastic constant $k_P$, a linear spring with elastic constant k in series with a dashpot with viscosity c, and a nonlinear spring with deformation-dependent coefficient $k_D$. Elastic and viscous properties are described by the linear spring and the series combination of spring and dashpot. The nonlinear spring is based on the so called densification concept, which occurs when the cells in a foam become closely packed at large compressions. More details about the model parameter will be given in section **4.6**.

As reported, the experimental data (circle symbols) in **Figure 4b** and **c** for cell spheroid deformation under a constant load were fitted with the following equation:

$$\sigma = [k_p + \gamma(1 - e^\varepsilon)^n \varepsilon]\varepsilon + \frac{c}{k}[k + k_p + \gamma(1 - e^\varepsilon)^n \varepsilon]\dot{\varepsilon} \tag{2}$$

derived from the Goga's phenomenological model[43] (blue line). The latter was able to describe not only the linear relationship between $\sigma$ and $\varepsilon$, but also the densification (non-linear) region. Residual standard deviations (RSD) were $2.7 \cdot 10^4$ and $1.1 \cdot 10^3$, for NIH/3T3 and PANC-1, respectively. By fitting the data of 15 cell spheroids for each cell line, the values of the parameters $k_p$, $\gamma$ and n shown in the table of **Figure 5b** were obtained.

The viscosity parameter c of the model was found by analysing the transient behaviour under compression. Images of the cell spheroids were taken as a function of time after applying the coverslip load, until a steady deformation was reached. As an example of the observed experimental trends, the transient deformation is plotted in **Figure 5** in the case of an NIH/3T3 spheroid (**c**) and a PANC-1 (**d**) spheroid. A plateau value was reached after *ca.* 100 s in the case of the NIH/3T3 spheroid, while for the PANC-1 spheroid a steep rise was followed by a continuous growth at a slower slope. In any event, the observed time scale to reach steady state (ca. 100 s) is well below the characteristic time for a significant cellular rearrangement within a spheroid. In the limit of vanishing deformation, the Goga's phenomenological model reduces to the equation:

$$\sigma_{\varepsilon \to 0} = \frac{c}{k}(k + k_p)\dot{\varepsilon} \tag{3}$$

from which the initial slope can be obtained as:

$$\dot{\varepsilon}(t = 0) = \frac{k\,\sigma}{c\,(k + k_p)} \tag{4}$$

By taking k ≫ $k_p$, which is typical for the compression of foams, the initial slope can be approximated as:

$$\dot{\varepsilon}(t = 0) \cong \frac{\sigma}{c} \quad (5)$$

The initial slope of the transient experimental data for a given applied stress can therefore be used to calculate the viscosity parameter c, which is also included in **Figure 5b** and has essentially the same value for the two cells lines.

## 3. Conclusions

In this work, a novel rheo-optical assay based on a compression test with microscopy glass coverslips as the applied load is proposed for the characterization of mechanical properties of cell spheroids in standard cell culture plates. The technique does not require costly instrumentation or specific laboratory skills and can be implemented in any biology lab equipped with cell culture facilities suitable to prepare cell spheroids. The latter can be transferred with a pipette to the wells of a standard cell culture plate and loaded by microscope coverslips placed in simple 3D printed inserts. Images of the compressed cell spheroids can be acquired by a routine inverted microscope or by a cell phone, both in transient and steady state conditions. Basic image analysis techniques can then be applied to measure the size and deformation of the compressed spheroid, thus allowing to calculate the stress and deformation values. The assay has been validated on an agarose gel particle by using independently measured values of the rheological parameters.

While the elastic modulus can be simply calculated as the slope of the linear region of the stress vs deformation curve, it should be pointed out that the same data can be used to calculate other rheological parameters by using models from the literature or developing new ones. As an example, we have used a phenomenological model originally derived to describe the mechanical behaviour of foams and showed how model parameters can be obtained by data fitting in transient and steady state regime. The model includes the so called densification process at high deformation, where the sample is compacted by the collapse of the cells of the foam. In the case of cell spheroid, volume changes due to loss of fluids occur at relatively large values of the imposed stress (about 15% at 15 kPa and 17% at 40 kPa for CT26 cells[28]), likely due to output of water from the ECM, which is estimated to take about 15% of the total volume. Hence, in the linear region used to calculate the elastic modulus the cell spheroid volume can be considered as essentially constant. The densification of the cell spheroids at higher deformation is included in the phenomenological foam model that we have used. Further investigation in this area can be carried out by applying our compression assay but is outside the scope of the present work.

The implementation of the compression assay in standard cell culture plates is one of its most effective novelties, not only because of its broad lab applicability, but also since it allows to test different conditions by well-established biological methods. Other methods of cell spheroids characterization from the literature, though quite accurate and well designed, are neither so user-friendly, due to the need of specialized equipment, nor suitable to biological tests on large scale. In addition, the assay can be applied to other samples, such as tissue biopsies. Hence, the potential impact of the proposed assay is quite extensive and can provide a boost to the current toolbox of mechanobiology research. For example, spheroid mechanical properties can be used as biomarkers for cancer diagnosis and prognosis, as changes in these properties have been linked to disease progression and to the impact of their sensitivity to anti-cancer treatments[44].

# 4. Materials and Methods

## 4.1. Preparation of agarose gel

Aqueous solutions of agarose (Agarose E, Condalab, #8100) at 2.5%wt, were prepared on a hot plate magnetic stirrer at 200°C for 20 minutes. Rhodamine B (Rhodamine B (C.I.45170) for microscopy 81-88-9) was added to the agarose solution for the sake of image clarity. As shown in **Figure 6**, 2 µL of aqueous solutions of agarose are quickly transferred by a pipette into a cell culture multi-well plate (Nunclon Δ Multidishes, 48 wells, flat bottom, Thermo Scientific, Nunc 150687) filled with 600 µL low viscosity silicone oil (Silicone oil 20 cSt 25°C, 63148-62-9, Sigma-Aldrich) at room temperature. The abrupt cooling of the agarose solutions results in the gelation of the latter in the shape of spherical particles suspended in the oil phase. The so obtained spherical gel particle is aspirated with a pipette and transferred to a 35 mm Petri dish filled with silicone oil. The load is applied by gently immersing a microscope coverslip (24mm x 24mm, thickness 0.13-0.16 mm, weight 0.21 g, Knittel glass n. VD12424Y1A.01) in the silicone oil centred with respect to the agarose gel particle (this procedure can be facilitated by drawing a centre target as the intersection of the two diagonals of the square coverslip with a glass marker and a ruler).

## 4.2. Cell culture and spheroids formation

One cancer cell line (PANC-1) and a non-tumour one (NIH/3T3) were used. All cells were cultured in their standard growth medium in 2D monolayers under the typical cell culture conditions, at 37°C in a humidified atmosphere (5% $CO_2$). NIH/3T3 mouse fibroblasts and PANC-1 human pancreatic carcinoma cells were cultured in Dulbecco's Modified Eagle's Medium (DMEM) supplemented with 10% (v/v) Fetal Bovine Serum (FBS), 1% (v/v) antibiotics (50 units/mL penicillin and 50 mg/mL streptomycin) and 1% (v/v) L-glutamine.

Spheroids were produced using the liquid overlay technique[7,24]. Specifically, 1% agarose solution was prepared by dissolving agarose powder (E AGAROSE, Conda, Cat nº 8100) in water at 200°C for 20' using a magnetic stirrer to homogenize the solution. Then, as shown in **Figure 7**, agarose solution was rapidly pipetted in 200 μl aliquots into the wells of a 48-well culture dish (Nunclon Δ Multidishes, 48 wells, flat bottom, Thermo Scientific) under sterile conditions and allowed to cool down. By capillarity, agarose solution rises along the walls of the wells, thus gelifying in a few minutes and forming a hemi-spherical meniscus. The non-adhesive concave surface promotes the collection of cells in the meniscus and cell-cell adhesion establishment; this results in the formation of cell aggregates and finally spheroids after an adequate incubation time, depending on cell type and concentration.

To generate spheroids, cells were harvested from monolayer cultures and counted. $8 \cdot 10^3$ and $3 \cdot 10^3$ NIH/3T3 and PANC-1 cells, respectively, were seeded in single wells pre-coated with non-adhesive agarose and covered with the cell growth medium. The multiwell plate was then incubated under typical cell culture conditions at 37 °C to allow spheroid formation. Typically, 5-10 days (depending on cell line) were required to obtain compact spheroids of adequate size. Minimum information of spheroids[45] for each cell lines were reported in **Supplementary Table 2** and **3** (NIH/3T3 and PANC-1, respectively).

### 4.3. Compression test and microscopy

The compression test was carried out by applying one or more microscope coverslips on a gel particle or cell spheroid. The positioning of the coverslips was facilitated by using a plastic frame (see Figure 1b) as a positioning guide. The frame was drawn by means of Autodesk Fusion 360 and 3D printed in PVA using Ender 3 S1 printer.

Spherical gel particle and spheroid morphological evolution, as a function of stress applied, was followed by Laser Scanning Confocal Microscope (LSM) 5 Pascal (Carl Zeiss Advanced Imaging Microscopy, Jena, Germany). Several independent fields of view were acquired by a high-resolution high-sensitivity monochromatic CCD video camera (Hamamatsu Orca AG, Japan) using a 5 x and 10x objectives.

### 4.4. Image Analysis

In order to quantify mechanical stimuli effect on spheroids evolution, morphological response of cellular spheroids, in terms of area and diameter (A and D, respectively), were measured by using an image analysis software (Image Pro Plus 6.0, Media Cybernetics). This measurement was made on the diametral plane for each image acquired.

Area evaluation plays a crucial role to quantify the applied stress; in details, the stress, $\sigma$, was defined as $(F_W - F_A)/A$, where $F_W$ was the glass coverslips weight, while $F_A$ was the Archimedes force, related to PBS in which spheroid is seeded. The applied stress, $\sigma$, was related to the local deformation, i.e., the strain, referred as $\varepsilon$. To define $\varepsilon$ as function of a measured parameter, the assumption of constant volume, before and after the compression, was made. In this way, the strain was evaluated as $\varepsilon = \delta/R$, i.e., as the normalised penetration under compression.

### 4.5. Rheological setup

Rheological measurements were performed by means of a stress controlled rheometer (Anton Paar Physica MCR 301 Instruments) equipped with parallel plates having diameters of 75 mm (PP75/TI/S-SN2005). Rheological tests were run at room temperature (23°C), temperature was controlled by a Peltier cooler/heater connected to a circulating water bath (Lauda) and let to equilibrate at the measuring value for 3 minutes before each test.

In order to polymerize the sample between plates, rheometer temperature was set to 70°C. Then ~10 mL of agarose solution was distributed on the bottom plate, and the upper plate lowered to a gap of 2 mm. To limit water evaporation, ~400 µL of oil (Wacker Wacker® AK 5 Silicone Fluid) were added on sample edge. Pre-shear was run to homogenise sample structure after loading, imposing a constant shear rate ($\dot{\gamma}$) of 200 1/s for 1 minute. Sample was cooled down to 23°C and let to complete polymerization for 30 minutes before running rheological characterization.

Strain sweep experiments were performed to identify the linear viscoelasticity range. Storage and loss moduli ($G'$ and $G''$, respectively) were measured at two fixed angular frequencies ($\omega$ = 10 and 1 rad/s) varying the strain in the range 0.01 – 10% and showed a linear behaviour for strains lower than ~ 0.5%. Frequency sweep was investigated at strain 0.05 % varying frequency in the range 30 – 0.1 rad/s.

Rheo-optical compression test was reproduced also under the rheometer by imposing a step by step incremental compression of the sample between the two plates by applying controlled values of normal force ($F_N$), in the range 10 - 40 N, with steps of 5 N. For each value, $F_N$ was kept constant for 10 minutes, while monitoring the value of sample gap (h). The compression stress applied was calculated as $\sigma = 4F_N/\pi D^2$, where D was the tool diameter, and related to the corresponding strain, calculated as $\varepsilon=(h_0-h)/h_0$, where $h_0$ is the initial value of gap (2 mm) and h is the current value of the gap, measured during the test.

### 4.6. Phenomenological model of foam compression

The phenomenological model used to fit the compression data is the one proposed by Goga[43]. As shown in **Figure 5a**, the model is based on the parallel combination of three elements: a linear spring with an elastic constant $k_P$, a linear spring with elastic constant k in series with a dashpot with viscosity c, and a nonlinear spring with deformation-dependent coefficient $k_D$. The equations describing the mechanical behaviour of the three elements are as follows:

$$\sigma_M = \sigma_K = \sigma_C \tag{6}$$

$$\varepsilon_M = \varepsilon_K + \varepsilon_C \tag{7}$$

$$\sigma = \sigma_M + \sigma_P + \sigma_D \tag{8}$$

$$\varepsilon = \varepsilon_M = \varepsilon_P = \varepsilon_D \tag{9}$$

$$\sigma_K = k\varepsilon_K \tag{10}$$

$$\sigma_C = c\dot{\varepsilon}_C \tag{11}$$

$$\sigma_P = k_P \varepsilon_P \tag{12}$$

$$\sigma_D = k_D \varepsilon_D = \gamma(1 - e^{\varepsilon_D})^n \varepsilon_D \tag{13}$$

$$\dot{\varepsilon} = \dot{\varepsilon}_M = \dot{\varepsilon}_K + \dot{\varepsilon}_C = \frac{\dot{\sigma}_M}{k} + \frac{\sigma_M}{c} \tag{14}$$

$$\sigma + \frac{c}{k}\dot{\sigma} = (k_P + k_D)\varepsilon + \frac{c}{k}(k + k_P + k_D)\dot{\varepsilon} \tag{15}$$

which then leads to **equation (2)** in the constant stress case investigated in this work.

### 4.7. Statistical analysis

At least 15 spheroids were analysed for each cell lines (tumoral and non-tumoral) for rheo-optical compression tests. Data are expressed as mean ± standard error of mean (SEM).

# Acknowledgements

Vincenzo Pepe contributed to the experiments, data analysis, and data interpretation as part of his master thesis. We would like to express our gratitude to Dr. Speranza Esposito and Dr. Valeria Rachela Villella for their invaluable support to the cell cultures.

## Author contributions

R.F. cultured cells, prepared spheroids and agarose particles, performed rheo-optical compression tests and rheological measurement, performed all live imaging experiments, analysed data and wrote the manuscript draft.

S.C. and S.G. designed and supervised the project, provided funds, wrote, reviewed and edited the manuscript.

All authors approved the manuscript.

## Competing interests

The authors declare no competing interests.

# Figure Legends

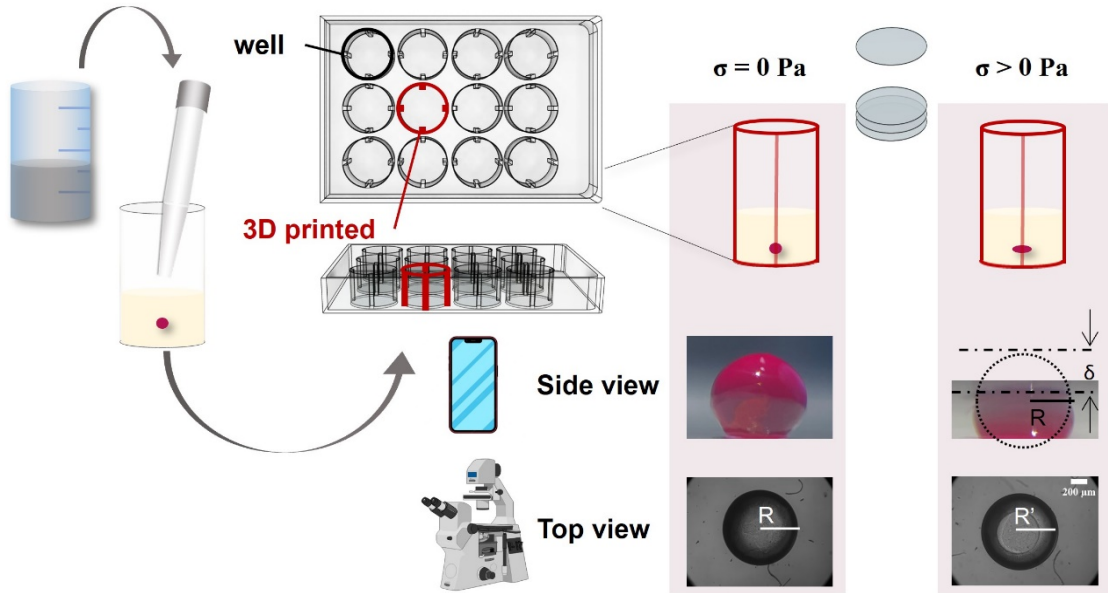

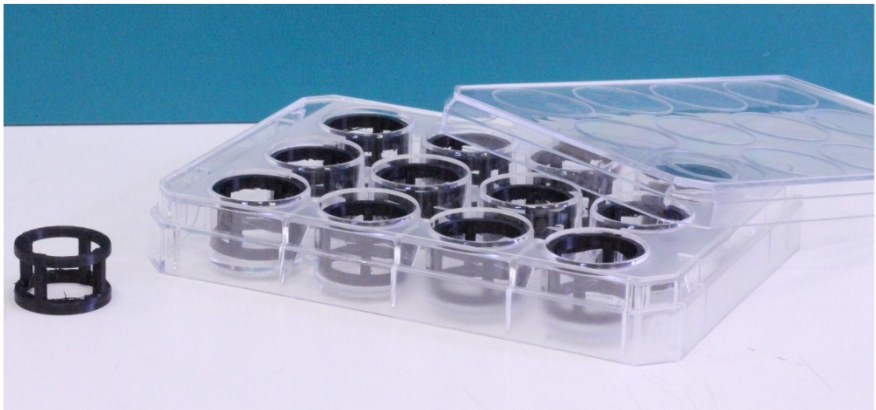

*Figure 1. a) Agarose particles preparation (see paragraph **4.1** for more details) and our rheo-optical compression assay scheme. Morphological response of an agarose particle, taken as a qualitative representation of a given experimental condition, at σ>0 Pa (second column), recorded by means of smartphone (first row) and microscope (second row), compared to the control condition (σ=0 Pa, first column). b) 3D printed coverslip guide (in black) for our rheo-optical compression assay.*

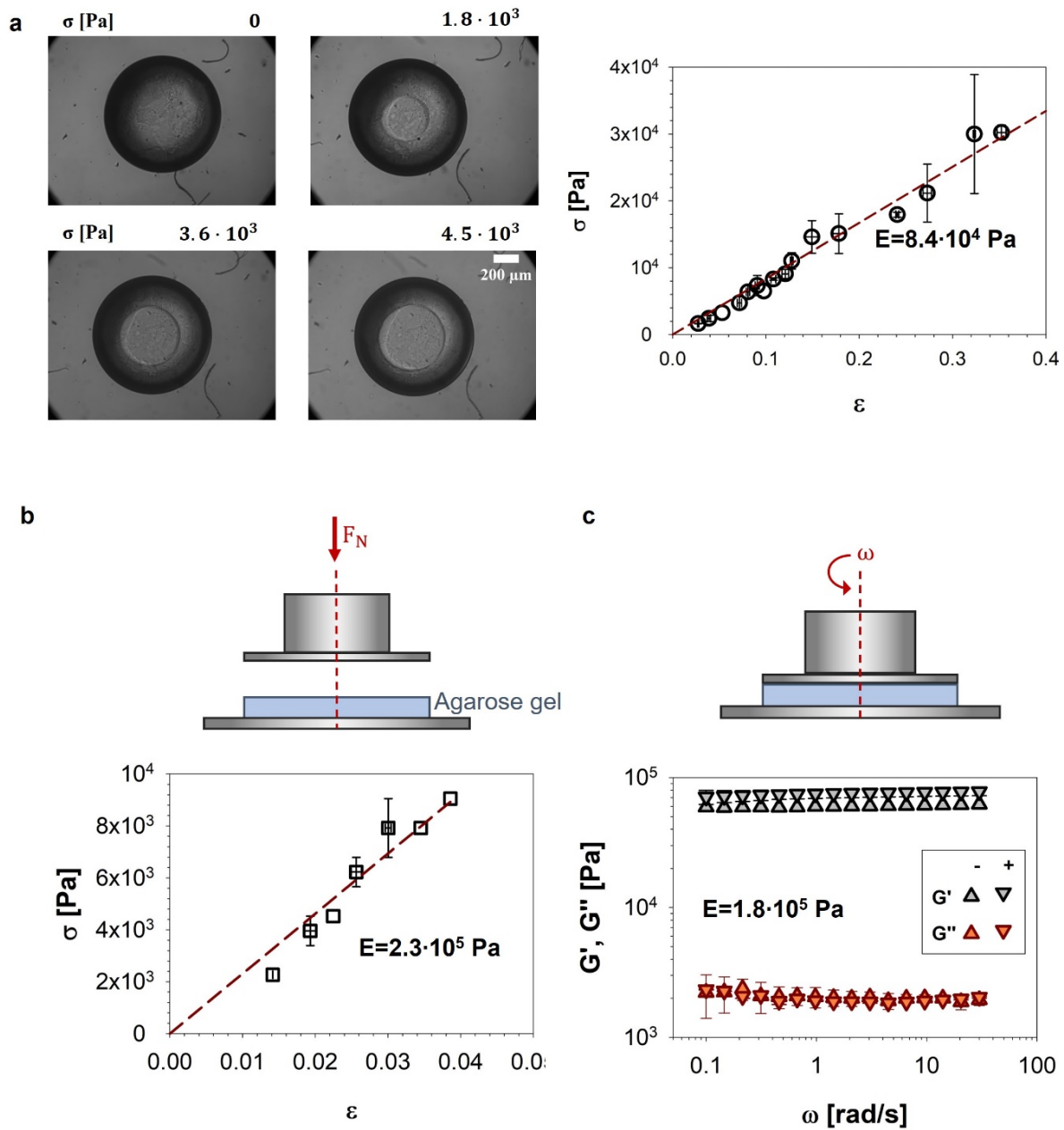

*Figure 2. a left) Representative phase-contrast microscopy images showing the morphological response of agarose particles at four different values of σ (from 0 to ~5·10³ Pa), corresponding to a different number of coverslip glasses. Scale bar: 200 μm. a right) Plot of the applied stress, σ, and strain, ε, in our rheo-optical compression test. The dashed line is a linear fit to the data. b) Plot of stress vs strain in compression tests performed by means of the rotational rheometer. The dashed line is a linear fit to the data. c) The shear storage modulus (G'(ω), ▲and ▽) and the shear loss modulus (G''(ω), ▲ and ▽)) versus the angular frequency (ω), obtained by oscillatory tests, for agarose particles, before (▲and ▲) and after (▽and ▽) the compression test performed in the same rotational rheometer. All the data are calculated as average ± standard deviation, over at least 2 independent measurements (overall 15 independent measures being available in the entire range investigated). The horizontal error bar is merely visible, being smaller than bullet size.*

**I) Spheroids preparation**    **II) Rheo-optical compression assay**

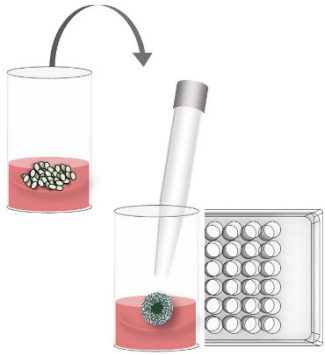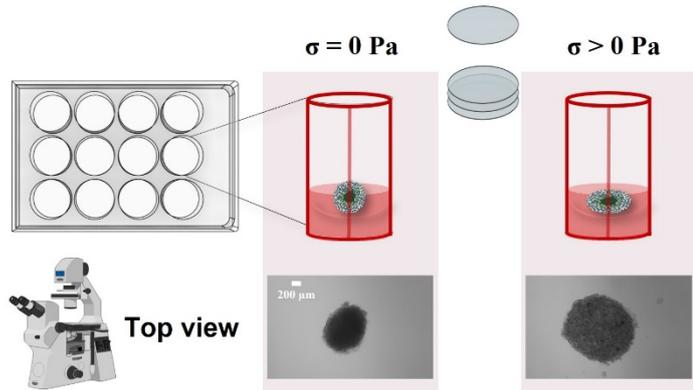

*Figure 3. a) Cellular spheroids preparation (see subsection **4.2** for more details) and b) a schematic of the rheo-optical compression assay. The optical microscopy images in the bottom of b) refer to a NIH/3T3 spheroid taken as an example to show the spheroid shape before (σ=0 Pa) and after the compression (σ>0 Pa) obtained by the application of glass coverslips.*

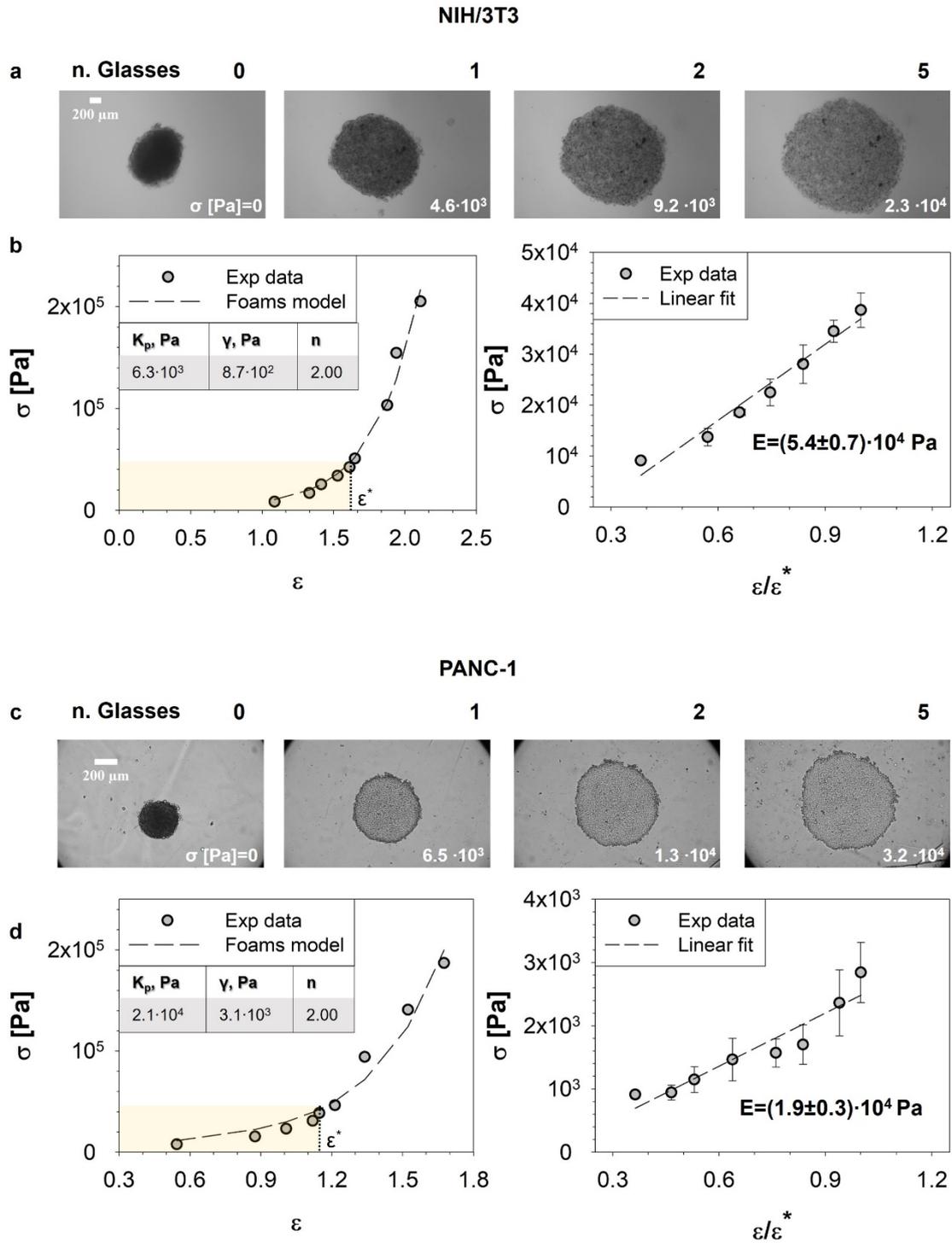

*Figure 4. (a and c) Representative phase-contrast microscopy images showing the morphological response of NIH/3T3 and PANC-1 spheroids at four different values of σ (from 0 to ~4·10⁴ Pa), corresponding to a different number of coverslip glasses. Scale bar: 200 μm. See **Supplementary Video 2 and 3**. (b and d, left) Stress-strain value quantified for a single representative spheroid of NIH/3T3 and PANC-1 cell line. The dashed curve is a fit of the foams model described in subsection 2.5. The rectangular region highlighted in yellow corresponds to the data (up to the value $\varepsilon^*$) used to calculate the Young's modulus. (b and d, right) The Young's modulus, denoted as E, quantified using the linear region of the stress-strain plot ($\varepsilon < \varepsilon^*$) for both NIH-3T3 and PANC-1 cell lines, respectively.*

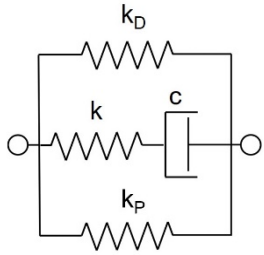

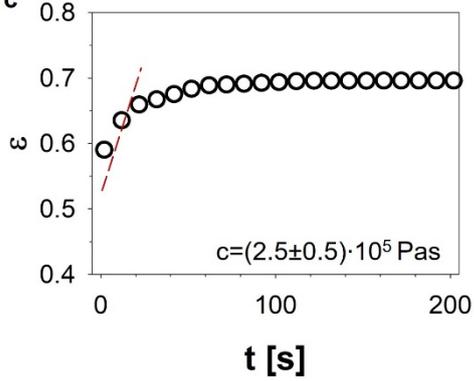
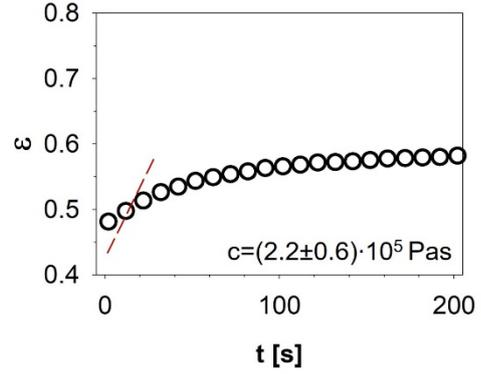

| | NIH/3T3 | | PANC-1 | |
|---|---|---|---|---|
| | Value | SEM | Value | SEM |
| $K_p$, Pa | $2.1 \cdot 10^3$ | $1.7 \cdot 10^2$ | $1.6 \cdot 10^4$ | $1.8 \cdot 10^3$ |
| $\gamma$, Pa | $3.2 \cdot 10^3$ | $2.0 \cdot 10^2$ | $7.7 \cdot 10^2$ | $4.2 \cdot 10^1$ |
| n | 2.0 | 0 | 2.0 | 0 |
| c, Pa·s | $2.5 \cdot 10^5$ | $4.8 \cdot 10^4$ | $2.3 \cdot 10^5$ | $5.7 \cdot 10^4$ |

*Figure 5. a) Schematic of the phenomenological model used to fit experimental data. b) Table with the model parameters obtained by fitting steady ($k_p$, $\gamma$ and n) and transient (c) data. c) and d) Transient strain $\varepsilon$ vs time for a NIH/3T3 and PANC-1 spheroid, respectively. The dashed lines represent the initial slope at t=0, which was used to calculate the value of c in the table of b).*

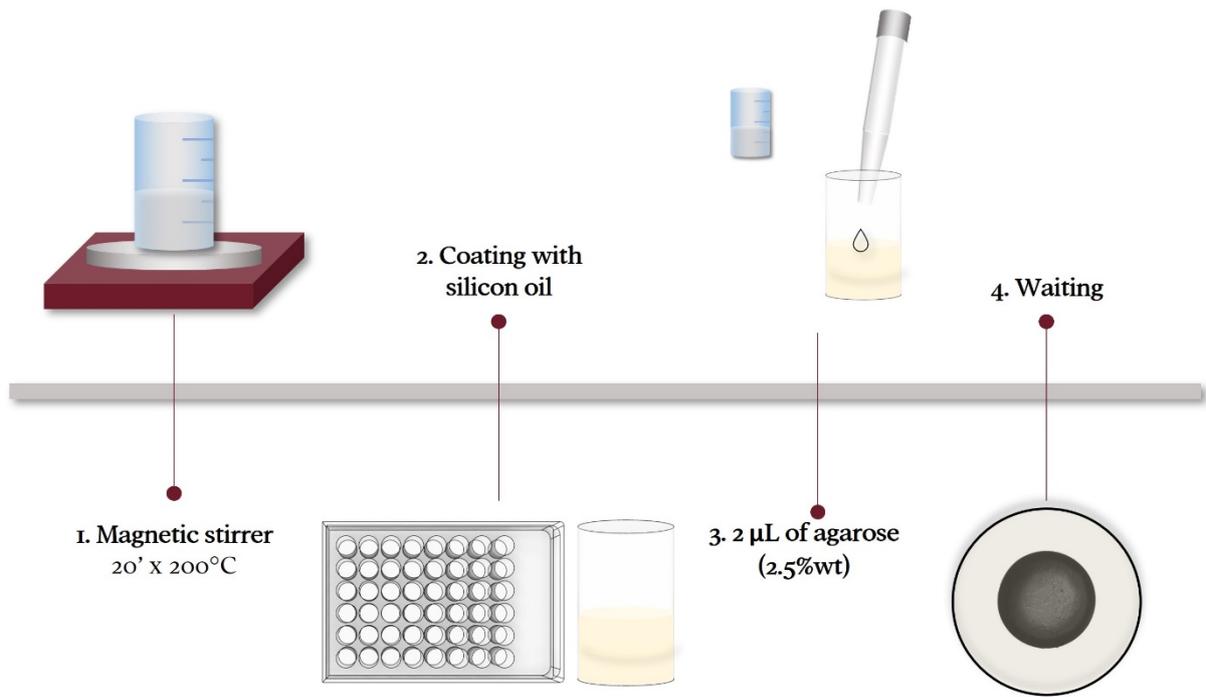

*Figure 6. A model of the agarose (2.5%wt) particles formation process. Particles formation can be divided into different stages: (1) aqueous solutions preparation on a hot plate magnetic stirrer at 200°C for 20 minutes; (2) transfer of agarose-aqueous solution by a pipette into a cell culture multi-well plate filled with 600 µL low viscosity silicone oil; (3) gelation and formation of compact particles by abrupt cooling.*

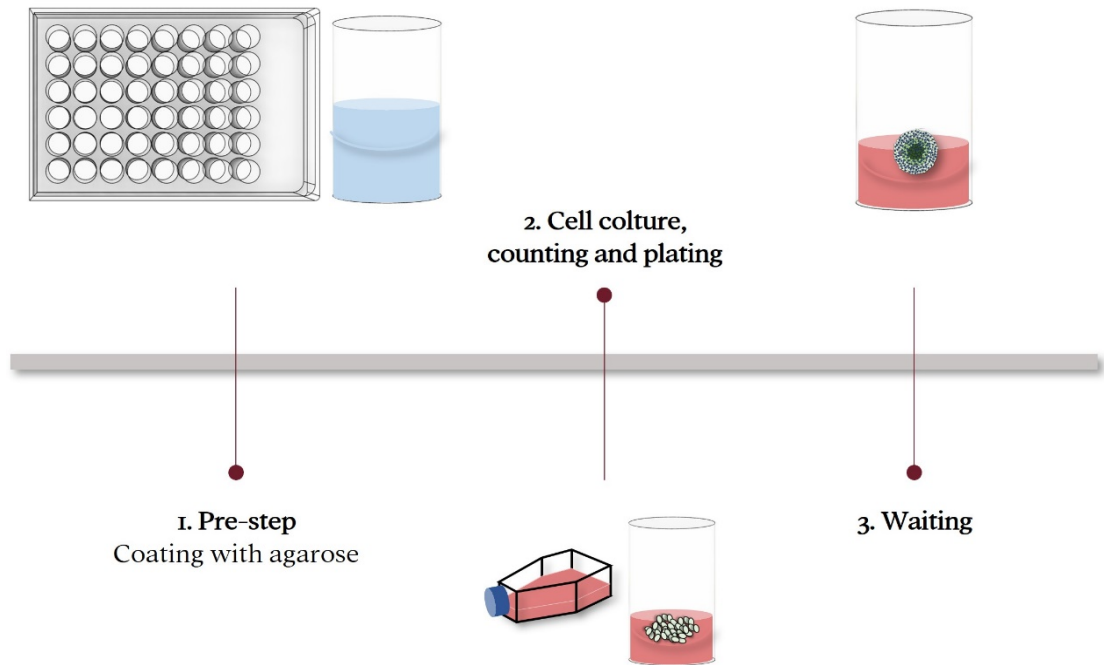

*Figure 7. A model of the spheroid formation process. Spheroid formation can be divided into different stages: (1) agarose (2%wt) solution was prepared and rapidly pipetted in 200 μl aliquots into the wells of a 48-well culture dish and allowed to cool down, obtaining a hemi-spherical meniscus by gelification; (2) cells, cultured in their standard growth medium in 2D monolayers, were counted and seeded in single wells pre-coated with non-adhesive agarose and covered with the cell growth medium; (3) formation of compact spheroids after 5-10 days (depending on cell line).*